\documentclass[lettersize,journal]{IEEEtran}
\usepackage{amsmath,amsfonts}
\usepackage{algorithmic}
\usepackage{algorithm}
\usepackage{array}
\usepackage[caption=false,font=normalsize,labelfont=sf,textfont=sf]{subfig}
\usepackage{textcomp}
\usepackage{stfloats}
\usepackage{url}
\usepackage{verbatim}
\usepackage{graphicx}
\usepackage{cite}
\usepackage{siunitx}
\usepackage{hyperref}
\usepackage{url}
\usepackage[skins]{tcolorbox}
\usepackage{multirow}
\usepackage{svg}
\usepackage{amsmath,amssymb}
\usepackage{enumitem}
\usepackage{float}
\usepackage{adjustbox}
\usepackage{soul}
\usepackage{xcolor}
\usepackage{ifthen}
\usepackage{booktabs}
\usepackage{pifont}

\newcommand{\dcircle}[1]{\ding{\numexpr171 + #1}}

\newboolean{showcomments}
\setboolean{showcomments}{true}
\ifthenelse{\boolean{showcomments}}
 {\newcommand{\mynote}[2]{
      \fbox{\bfseries\sffamily\scriptsize#1}
        {\small$\blacktriangleright$\textsf{\emph{#2}}$\blacktriangleleft$}}}
        { \newcommand{\mynote}[2]{}}

\usepackage{xspace}
\newcommand{\toolname}{{DexBERT}\xspace}

\usepackage{relsize}

\begin{document}

\title{{\relscale{0.9} \toolname: Effective, Task-Agnostic and Fine-grained Representation Learning of Android Bytecode}}

\author{Tiezhu~Sun,
        Kevin~Allix,~\IEEEmembership{}
        Kisub~Kim\textsuperscript{*},~\IEEEmembership{}
        Xin~Zhou,~\IEEEmembership{}
        Dongsun~Kim,~\IEEEmembership{}
        David~Lo,~\IEEEmembership{Fellow, IEEE},
        Tegawendé~F.~Bissyandé,~\IEEEmembership{Member, IEEE}
        and~Jacques~Klein,~\IEEEmembership{Member, IEEE}
\IEEEcompsocitemizethanks{\IEEEcompsocthanksitem T. Sun, K. Allix, T. Bissyandé and J. Klein are with the University of Luxembourg.\protect\\
E-mail: \{tiezhu.sun,kevin.allix,tegawende.bissyande,jacques.klein\}@uni.lu
\IEEEcompsocthanksitem \textsuperscript{*}K. Kim is the corresponding author and is with Singapore Management University along with X. Zhou and D. Lo. \protect \\
E-mail: \{kisubkim,davidlo\}@smu.edu.sg,xinzhou.2020@phdcs.smu.edu.sg
\IEEEcompsocthanksitem Dongsun~Kim is with Kyungpook National University. \protect\\
E-mail: darkrsw@gmail.com}
\thanks{Manuscript received December 2, 2022; accepted August 24, 2023.}}

\markboth{Journal of \LaTeX\ Class Files,~Vol.~xx, No.~x, xx~xxxx}%
{Shell \MakeLowercase{\textit{et al.}}: A Sample Article Using IEEEtran.cls for IEEE Journals}


\maketitle

\begin{abstract}
The automation of an increasingly large number of software engineering tasks is becoming possible thanks to Machine Learning (ML). One foundational building block in the application of ML to software artifacts is the \emph{representation} of these artifacts (\textit{e.g.}, source code or executable code) into a form that is suitable for learning. 
Traditionally, researchers and practitioners have relied on manually selected features, based on expert knowledge, for the task at hand.
Such knowledge is sometimes imprecise and generally incomplete. 
To overcome this limitation, many studies have leveraged representation learning, delegating to ML itself the job of automatically devising suitable representations and selections of the most relevant features. 
Yet, in the context of Android problems, existing models are either limited to coarse-grained whole-app level (\textit{e.g.}, \texttt{apk2vec}) or conducted for one specific downstream task (\textit{e.g.}, \texttt{smali2vec}).
Thus, the produced representation may turn out to be unsuitable for fine-grained tasks or cannot generalize beyond the task that they have been trained on. 
Our work is part of a new line of research that investigates effective, task-agnostic, and fine-grained universal representations of bytecode to mitigate both of these two limitations. Such representations aim to capture information relevant to various low-level downstream tasks (\textit{e.g.}, at the class-level). We are inspired by the field of Natural Language Processing, where the problem of universal representation was addressed by building Universal Language Models, such as BERT, whose goal is to capture abstract semantic information about sentences, in a way that is reusable for a variety of tasks. 
We propose \toolname, a BERT-like Language Model dedicated to representing chunks of DEX bytecode, the main binary format used in Android applications. 
We empirically assess whether \toolname is able to model the DEX \emph{language} and evaluate the suitability of our model in three distinct class-level software engineering tasks: Malicious Code Localization, Defect Prediction, {and Component Type Classification}. 
We also experiment with strategies to deal with the problem of catering to apps having vastly different sizes, and we demonstrate one example of using our technique to investigate what information is relevant to a given task.

\end{abstract}



\section{Introduction}
\label{sec:introduction}

\IEEEPARstart{P}{re-trained} models yielding general-purpose embeddings have been a recent highlight in AI advances, notably in the research and practice of Natural Language Processing (\textit{e.g.}, with BERT~\cite{bert}). 
Building on these ideas, the programming language and software engineering communities have attempted similar ideas of learning vector representations for code (\textit{e.g.}, \textsc{code2vec}~\cite{code2vec}) and other programming artifacts (\textit{e.g.}, bug reports~\cite{mani2019deeptriage}). 
Unfortunately, these pre-trained models for code embedding often do not generalize beyond the task they have been trained on~\cite{ASE2019-code2vecGeneralizability}.



In the Android research landscape, 
{many techniques have been proposed to address app classification problems~\cite{canfora2015detecting, hou2016droiddelver, singh2017dynamic, xiao2017back, xiao2016identifying, xu2018cdgdroid, ma2020droidetec,arslan2021androanalyzer}, most of which, however, can only handle coarse-grained tasks (\textit{i.e.}, at the whole-app level). 
Although there are a few works~\cite{narayanan2018multi,dong2018defect} targeting some fine-grained tasks at the class level, their learned representations have also not been shown to generalize to other class-level tasks.}

{Therefore, despite the good performance exhibited by existing approaches, there is still a research gap to be filled with the investigation of {\bf simultaneously fine-grained and task-agnostic} representation learning for Android applications.}
Indeed, advances in this direction will help researchers and practitioners who are conducting class-level tasks, such as malicious code localization or app defect prediction, to achieve state-of-the-art performance while reducing costs due to manual feature engineering or repetitive pre-training computations for multiple representation models.

Building a fine-grained task-agnostic model is, however, challenging since it essentially requires to capture knowledge relevant to a variety of tasks altogether and {at the low granularity of the representation}.
A few studies have investigated and built task-agnostic models in the field of software engineering. CodeBERT~\cite{CodeBERT} and \texttt{apk2vec}~\cite{narayanan2018apk2vec} are key representatives of these models.
On the one hand, while CodeBERT brings significant improvements, it cannot be directly used for representing Android apps for two main reasons: lack of source code in apps and limit of input elements.
{Even though it is technically feasible to apply it to the assembly language \texttt{Smali}, the performance is unsatisfactory, as evidenced by our experiments in Section~\ref{sec:bert-like_comparison}.}
On the other hand, \texttt{apk2vec} successfully constructs the behavior profiles of apps and achieves significant accuracy improvements while maintaining comparable efficiency. However, 
despite \texttt{apk2vec}'s success in app representation, its granularity, and graph-based design still have limitations~\cite{giger2012method,singh2015detection,tantithamthavorn2018impact,zhang2019finelocator,frick2020understanding}. Notably,
{\texttt{apk2vec} is designed to handle only app-level tasks, while our proposed approach is targeted at fine-grained tasks at the class-level}.


Towards addressing limitations of existing representation techniques (\textit{i.e.}, lack of a universal model for Android bytecode {at low granularity}), we propose in this paper \toolname, a {fine-grained} and task-agnostic representation model for the bytecode in Android app packages.
The result of \toolname can be applied across various {class-level} downstream tasks (\textit{e.g.}, malicious code localization, defect prediction, \textit{etc.}).
Our approach first extracts features from \texttt{Smali} instructions (\textit{i.e.}, an assembly language for the \emph{Dalvik} bytecode used by Android’s Dalvik virtual machine.). 
It then combines embeddings of code fragments to build a model that can {address various class-level problems}.
Our pre-trained model can capture the essential features and knowledge by first learning an accurate general model of Android apps' bytecode.
The proposed aggregation methods allow \toolname to handle fine-grained {class-level} tasks, making \toolname capable of operating on lower granularity of Android artifacts than other state-of-the-art representation models.

To evaluate the effectiveness of \toolname, we first conduct a preliminary experiment to observe the feasibility of building a general model of Android app code. 
This experiment mainly focuses on pre-training BERT on \texttt{Smali} instructions to ensure that the generated embeddings contain meaningful features for various tasks. 
We observed clearly converging loss curves on all the pre-training tasks, with the pre-trained model achieving 95.30\% and 99.35\% accuracy on the masked language model and next sentence prediction tasks, respectively. 
Such performances demonstrate that our model indeed learned meaningful features and can be generalized to a variety of different tasks. 

We further perform a comprehensive empirical evaluation to measure the overall performance of \toolname on three class-level downstream tasks: Android malicious code localization, Android defect detection, {and Component Type Classification}. 
Android malicious code localization is a task whose goal is to identify the malicious parts of Android apps. 
Android defect detection locates defective code to help developers improve the security and robustness of Android apps. 
{Component Type Classification is a multi-class classification problem that has a distinct character compared to the two tasks mentioned above. This classification problem is introduced to provide a more comprehensive evaluation of \toolname's universality.}
Our experimental results show that \toolname can localize malicious code and detect defects with significant improvement over current state-of-the-art approaches (74.93 and 6.33 percentage points improvement for malicious code localization and defect prediction, respectively) in terms of accuracy.
{\toolname also significantly outperform other BERT-like baselines on the task of Component Type Classification by a roughly 20 percentage point increase in terms of F1 Score.}


The contributions of our study are as follows: 
\begin{itemize}
    \item We propose a novel BERT-based pre-trained representation learning model for Android bytecode representation, named \toolname. It can be used directly on various {\underline{class-level} (\textit{i.e.}, finer-grained than existing \underline{app-level} approaches)} downstream analysis tasks by freezing parameters of the pre-trained representation model when tuning the prediction model for a specific downstream task. 
    \item We propose aggregation techniques that overcome the limitations with the size of the input of BERT.
    \item We conduct a comprehensive evaluation that shows that \toolname achieves promising performance on multiple pre-training tasks and class-level Android downstream tasks.
    \item To ease replication, we share the dataset and source code to the community at the following address:\\
        \url{https://github.com/Trustworthy-Software/DexBERT}. 
\end{itemize}

\begin{figure*}[htp]
\centering
\includegraphics[keepaspectratio,width=0.95\linewidth]{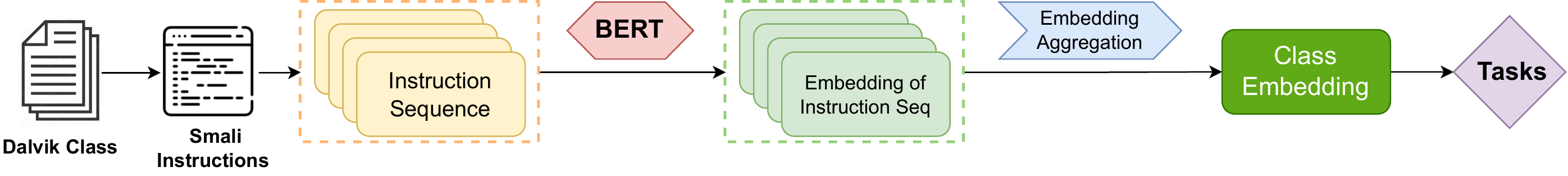}
\vspace{-3mm}
\caption{Overview of a class embedding by  \toolname.}
\label{fig:dexbert}
\vspace{-4mm}
\end{figure*}

\section{Background \& Motivation}
\label{sec:background}


\subsection{Representational Model}
\label{sec:back_bert}

Our approach, \toolname, aims at building a task-agnostic universal representation model for Android apps at low granularity.
Universal representation models are trained on large corpora to automatically capture general features that are relevant to various downstream tasks without training a model from scratch for each task.
For example, BERT~\cite{bert} has revolutionized the representation problem of Natural Language Processing (NLP). 
One advantage of BERT-like models is the strong separation between the generic representation and any specific task.
To build a language model, BERT pre-training relies on two fundamental tasks, Masked Language Model (MLM) and Next Sentence Prediction (NSP), without manual labels.
MLM forces BERT to capture relationships between words, and NSP forces BERT to capture relationships between sentences.
With such a mechanism, BERT can generate meaningful embeddings of the input sentences.
Universal models can also be useful in the field of software engineering, where many tasks that take software artifacts as input are investigated.
CodeBERT~\cite{CodeBERT} is one of the successful universal models that is used by a number of studies~\cite{9463106,app11114793,yuan2022deep,fujimori2022using} in various software engineering domains with promising performance.




As the purpose of our approach, \toolname, is to represent the bytecode in Android apps with concise representations, such as embedding vectors; we leverage and adapt the idea and architecture of BERT~\cite{bert} as well.
While BERT, coming from NLP, works with \emph{words} and \emph{sequences of words} (\textit{i.e.}, sentences or paragraphs), we disassemble Android apps into Smali code\footnote{Smali is a popular disassembler for Android applications (\url{https://github.com/JesusFreke/smali}).}. 
We then regard the Smali code as a flow of tokens that can be fed to a BERT-like model.

\subsection{Downstream Tasks}

Once a BERT-like model is pre-trained, it is able to output a representation of its input. 
This representation can then be used as an input to other models that can be trained (fine-tuning in BERT terminology) for a specific downstream task. 
We evaluate our approach on three class-level Android-specific tasks: malicious code localization, defect prediction, {and Component Type Classification}

Precise malicious code localization not only helps to assess the trustworthiness of existing app-level malware detection methods but also enables important applications such as studying malware behavior and engineering malware signatures.
Malicious code localization becomes particularly useful in the case of repackaged malware, as the major portion of apps' code remains benign, with only a small portion relevant to the attack~\cite{narayanan2018multi}. 

Software defects are errors or bugs built into the software due to programmers' mistakes, such as memory overflows and run-time exceptions~\cite{malhotra2016empirical}, during the software design or development process. 
These defects can raise serious reliability and security concerns. 
Automatically finding such defects is thus an important and active domain of research~\cite{dong2018defect}.

{As an additional measure to assess DexBERT's wide-ranging applicability, we introduced Component Type Classification, a third task at the class level. This task, which involves multi-class classification, serves as a contrast to the preceding two tasks that focused on security-related binary classification, and aims to provide a thorough evaluation of DexBERT's universality in different contexts.}

In the Android context, these three tasks require sub-app level representations (\textit{i.e.}, class-level) instead of whole-app level representations (\textit{e.g.}, apk2vec~\cite{narayanan2018apk2vec}) since the tasks pinpoint a specific location in an app.
Note that whole-app level representations transform an app into an embedding vector rather than transforming each element (\textit{e.g.}, class) of an app. Our approach can take a subset of bytecode in an app, and thus it can represent classes  as embedding vectors.
Theoretically, other class-level tasks beyond the three we identified are expected to benefit from \toolname as well, as long as the corresponding datasets are well-labeled and publicly available.


\subsection{Motivation}

As the population of Android apps is rapidly growing larger and the number of relevant issues (\textit{e.g.}, productivity and security problems) is increasing quickly, it is necessary to efficiently address the issues of the Android ecosystem; for example, one solution to multiple problems. 
However, most state-of-the-art approaches focus on building a model for a particular issue, such as Malware Detection (\textit{e.g.}, Drebin~\cite{arp2014drebin} and DexRay~\cite{daoudi2021dexray}) or Malicious Code Localization (\textit{e.g.}, MKLDroid ~\cite{narayanan2018multi} and Droidetec~\cite{ma2020droidetec}), or Software Defect Prediction (\textit{e.g.}, smali2vec~\cite{dong2018defect} and SLDeep~\cite{majd2020sldeep}). 

Instead of devising a single individual solution for each of those issues, there is value in investigating the possibility of having one single universal model that is able to capture and represent the relevant features and properties of application's code, which could be leveraged for a variety of tasks.
Some existing representation models generalize to a variety of problems, but target either a low granularity with source code (\textit{e.g.}, CodeBERT~\cite{CodeBERT}) or entire Android applications (\textit{e.g.}, apk2vec~\cite{narayanan2018apk2vec}).
CodeBERT requires native source code (\textit{i.e.}, written in programming languages), which is often not available for Android apps.
While decompilation of Java code is an option, it is, however not always complete and is often significantly different from real source code.
Moreover, the design of CodeBERT does not overcome the limitation with the number of input elements (512 for BERT) for Android apps (\textit{i.e.}, apps often contain more than millions of tokens). 
In addition, most existing Android representation approaches target the whole-app level (\textit{e.g.}, apk2vec~\cite{narayanan2018apk2vec}), which makes it difficult to explore fine-grained details of the problems.

Therefore, it would be a significant step forward if a representation model can take bytecode of apps and support various tasks at fine-grained levels.
The approach we propose, \toolname, does not require source code and is validated to cater to class-level tasks.
Meanwhile, we design an aggregation method to overcome the limitation of the 512 input elements in BERT-like models.
For \toolname users, they don’t need to pre-train again and only need to use the provided model to generate features for their own APKs. Then, they can do any class-level tasks they want.
Therefore, its {\bf reusability} can avoid training individual models for different tasks and save a large amount of time and effort.

\begin{figure*}[htp]
\centering
\includegraphics[width=\linewidth]{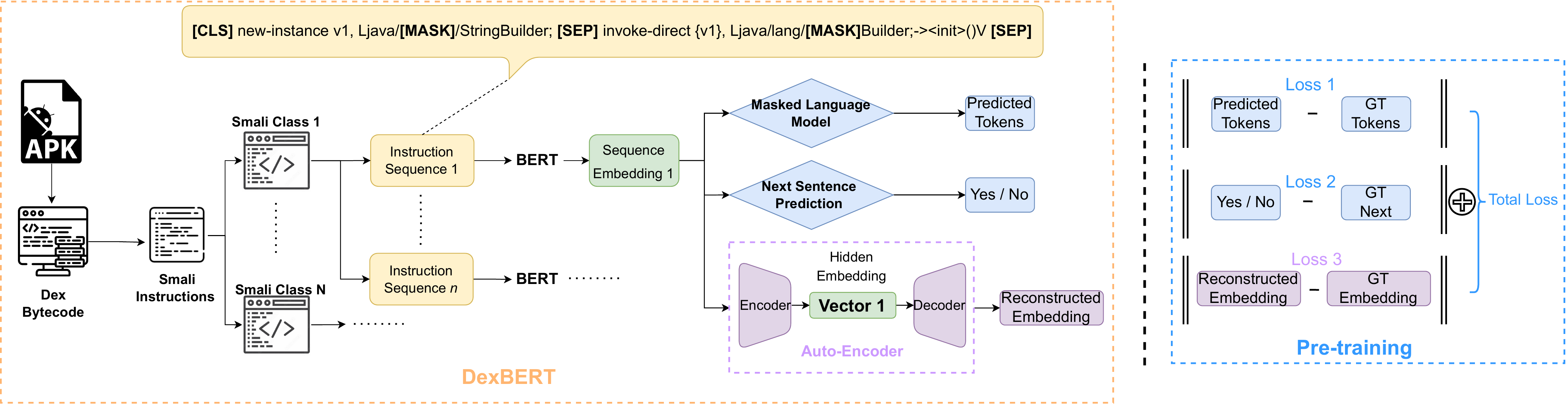}
\vspace{-5mm}
\caption{Illustrations of \toolname and Pre-training Loss Function. ``GT'' is an abbreviation for ``ground-truth''.}
\label{fig:detail_androbert}
\vspace{-3mm}
\end{figure*}

\section{Approach}
\label{sec:approach}


In this section, we first present the overview of \toolname workflow in Section~\ref{sec:overview}. Then, we illustrate details of \toolname in Section~\ref{sec:dexbert}, and we describe the applications of representations learned by \toolname on downstream tasks in Section~\ref{sec:prediction}.

\subsection{Overview}
\label{sec:overview}




\toolname focuses on extracting features from Android application Dalvik bytecode and targets fine-grained Android tasks (\textit{i.e.}, at class-level).
Our approach is clearly different from \texttt{apk2vec}~\cite{narayanan2018apk2vec}, which is a static analysis based multi-view graph embedding framework for app-level Android tasks, and CodeBERT, which is a bimodal pre-trained model for programming language (PL) and natural language (NL) tasks.
Specifically, \toolname takes disassembled \texttt{Smali} code from Dalvik bytecode as input and learns to extract corresponding representations (\textit{e.g.}, the embedding of a class). 
\texttt{Smali} is a text representation of Dalvik bytecode, in the same way, that assembly code is a text representation of compiled code.
Because \toolname supports various class-level tasks (while the original BERT is limited to relatively small inputs), there is a need to combine, or aggregate the representations of several sequences of instructions, or \emph{chunks}, into one single representation that covers the chosen input. 

As shown in Figure~\ref{fig:dexbert}, representation (or embedding) of the \texttt{Smali} Bytecode is learned by BERT during the pre-training phase.
This learned embedding can then be applied to class-level Android downstream tasks.
Specifically, we can easily extract the Bytecode of an Android application.
After disassembling each Dalvik class, we obtain the \texttt{Smali} instructions. 
This flow in \texttt{Smali} instructions (grouped in chunks) is fed to the BERT model in order to pre-train it, \textit{i.e.}, to learn how to represent \texttt{Smali} code.
Then, to obtain  the embedding of a \texttt{Smali} class, we aggregate the learned \toolname representation of each \texttt{Smali} instruction sequence present in the class.

\subsection{DexBERT}
\label{sec:dexbert}

We introduce the pipeline of DexBERT in Section~\ref{sec:bert_formulate}.
The mechanism of \toolname pre-training is presented in Section~\ref{sec:pretraining}.
In Section~\ref{sec:ae}, we introduce the Auto-Encoder, which is designed to reduce the dimensionality of the learned representation while keeping the key information.

\subsubsection{DexBERT}
\label{sec:bert_formulate}

As an assembly language, the \texttt{Smali} code disassembled from \emph{Dalvik} bytecode is a sequence of instructions. 
Similar to the original BERT, we pre-train our model on multiple pre-training tasks to force \toolname to capture general-purpose representations that can be used in various downstream tasks.

Specifically, as shown in the left box in Figure~\ref{fig:detail_androbert}, regarding each \texttt{Smali} instruction in each class as a text snippet we are able to create instruction pairs (similar to sentence pairs in original BERT) as input sequences.
In each pair, two instructions are separated by a reserved token \texttt{[SEP]}, and several randomly selected tokens (or "words") are masked.
For the masked language model, one of the two pre-training tasks of the original BERT, the goal is to correctly predict the tokens that are masked. 
The second original BERT pre-training task, Next Sentence Prediction, is turned here into Next Instruction Prediction, leveraging the pairs of instructions. In essence, this task's goal is to predict, given a pair of instructions, whether or not the second instruction follows the first one.

\subsubsection{Pre-training}
\label{sec:pretraining}
Pre-training plays a vital role in helping \toolname learn to generate meaningful embeddings. 
To ensure that the learned embeddings have general-purpose, the pre-training is supposed to be performed on multiple tasks simultaneously.
As described, we adopt and adapt the two pre-training tasks of the original BERT: \dcircle{1} masked language model (\textit{i.e.}, masked words prediction) and \dcircle{2} next sentence prediction as the main pre-training tasks.

In each pre-training iteration, the input sequence of instructions is fed into the BERT model, which generates a corresponding sequence embedding (as shown in the left box in Figure~\ref{fig:detail_androbert}).
The sequence embedding is then taken as input for the pre-training tasks.
Each task is a simple neural network with a single fully connected layer.
As shown in the right box in Figure~\ref{fig:detail_androbert}, a loss value is then calculated by comparing the output of each task head to the automatically created ground-truth (\textit{i.e.}, randomly masked tokens or binary label indicating whether the second statement indeed follows the first one or not).
The model weights of connections between neurons are adjusted to minimize the total loss value (\textit{i.e.}, the sum of all loss values for pre-training tasks) based on the back-propagation algorithm~\cite{rumelhart1986learning}.
Note that the pre-training tasks are only designed to help the BERT model learn meaningful features of input sequences and have little practical use in the real world.

\subsubsection{Auto-Encoder}
\label{sec:ae}
Even though the two aforementioned pre-training tasks could work well on \texttt{Smali} instructions, the dimensionality (\textit{i.e.}, $512\times768$, which is defined as the multiplication of token number in each input sequence $N$ and the dimension of the learned embedding vector $H$) of the generated representation for each sequence is quite large. 
Since there are usually hundreds or thousands of statements in a \texttt{Smali} class that need to be embedded, the dimensionality of the learned embeddings should be reduced before deployment to downstream tasks while preserving their key information. 
{In common practice for BERT-like models, the first state vector (of size 768) of the learned embedding is often used for this purpose.
In our case,} we add a third pre-training task, an Auto-Encoder, whose goal is to find a smaller, more efficient representation.

{A smaller representation is necessary primarily because APKs consist of a significantly larger number of tokens compared to typical textual documents and code files.
We provide a comparative analysis on token sequence length across three different data formats - textual documents (Paired CMU Book Summary \cite{bamman2013new}), code files (Devign \cite{zhou2019devign}), and APKs (DexRay \cite{daoudi2021dexray}), the number of BERT tokens in each dataset, denoted as [Mean]±[Deviation].
Specifically, the Paired CMU Book Summary has 1148.62±933.97 tokens, the Devign dataset contains 615.46±\num{41917.54} tokens, while the DexRay dataset contains a considerable 929.39K±11.50M tokens. This substantial quantity difference in tokens for APKs necessitates the effort to achieve as compact an embedding as possible for a given token sequence of Smali instructions.}

AutoEncoder~\cite{goodfellow2016deep} is an artificial neural network that can learn efficient small-sized embedding of unlabeled data.
It is typically used for dimensionality reduction by training the network to ignore the "noise".
The basic architecture of an Auto-Encoder usually consists of an Encoder for embedding learning and a Decoder for input regenerating (as shown in the left box in Figure~\ref{fig:detail_androbert}).
During the training process, the embedding is validated and refined by attempting to regenerate the input from the embedding, essentially trying to build the smallest complete representation of its input.

\begin{figure*}[htp]
\centering
\includegraphics[width=\linewidth]{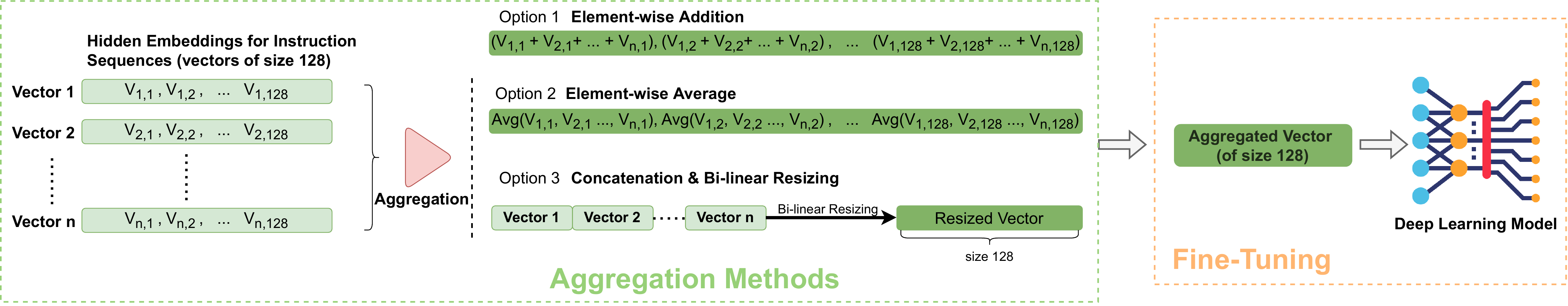}
\vspace{-3mm}
\caption{Illustrations of Different Embedding Aggregation Methods and Fine-tuning of Downstream Tasks.}
\label{fig:aggregation}
\vspace{-3mm}
\end{figure*}

The encoder in our approach consists of two fully connected layers with 512 and 128 neurons, respectively.
Symmetrically, the decoder consists of two fully connected layers with 128 and 512 neurons, respectively.
The sequence embedding (with the size of $512\times768$) learned from BERT is both the input of the Encoder and the output target of the Decoder in the pre-training process.
After comparative experiments, we opted for a hidden embedding with size $128$ as the final representation of the original instruction sequence.
{We provide an analysis of different embedding sizes in Section~\ref{sec:AE_Ablation}.}
Compared with the raw BERT embedding, the dimension of the final sequence embedding is \num{3072} times smaller, from $512\times768$ to $128$.
Consequently, in Figure~\ref{fig:detail_androbert}, the size of \emph{Vector 1}, which is the  embedding of \emph{Instruction Sequence 1} yielded by the Encoder, is 128.

\subsection{Class-level Prediction Model}
\label{sec:prediction}
In order to validate the effectiveness of learned \toolname representation, we apply it to three class-level Android analysis tasks.
To perform these class-level tasks, we need an efficient representation for each class, and we thus need a method to aggregate the embeddings of the instruction sequences of each class into one single embedding. We introduce this method in  Section~\ref{sec:aggregation}.
We then present the details of the prediction model in Section~\ref{sec:pre_model}.

\subsubsection{Aggregation of Instruction Embedding}
\label{sec:aggregation}

The representations learned from \toolname are for instruction sequences.
Usually, each \texttt{Smali} class consists of many methods, each of which contains a certain, often large, number of statements.
The number of sequence vectors in each class is indeterminate, while the shape of the input to the class-level prediction model (neural network) is supposed to be fixed. 
Specifically, within each class, there are many sequence embedding vectors with a size of 128, which are expected to be combined into one single vector.
Thus, embedding aggregation is required to obtain the final class-level representation.

{We primarily have {three} reasons to solve the long instruction sequence problem by splitting a \texttt{Smali} class into snippets and aggregate the learned snippet embedding into one class embedding. 
First, \texttt{Smali} instructions are individual commands performing specific operations in the Android run-time environment. 
Although they often appear in sequences, each \texttt{Smali} instruction operates largely independently, not necessarily bearing the kind of interdependence seen in words within natural language sentences. 
Second, the class representation, which aggregates the instruction embeddings, retains a sense of the overall structure and function of the class while being context-aware.}
{Last, while there might be some loss of context when splitting long sequences, we believe the trade-off between computational efficiency and a minor potential loss of context is justified.
To give a quantified measure of this trade-off, let's consider the GPU memory required. Doubling the input length limit for a BERT model would necessitate four times the initial GPU memory, a demand that most standard devices cannot meet, particularly for even longer sequences.
However, the high memory cost can be avoided without significant performance loss if we perform an average of 2.69 splits on long sequences in the adopted datasets, as shown in our experiments.
}

While it would be possible to leverage another step of representation learning to devise a strategy to aggregate several representations, we instead opt for less computationally expensive approaches. 
In order to adapt to the variability of vector numbers in different \texttt{Smali} classes, as shown in Figure~\ref{fig:aggregation}, we propose to aggregate these sequence embedding vectors of size 128 into one single vector of size 128 by performing simple element-wise average, element-wise addition, {or random selection}.
Another method we investigate is a general approach to vector reshaping in computer vision: concatenation \& bilinear resizing. 
Specifically, we first concatenate the embedding vectors into a long vector and resize it with the bilinear interpolation algorithm~\cite{fadnavis2014image}.
We investigate these four methods in Section~\ref{sec:rq_aggregate} and show that these simple methods are effective.

\subsubsection{Prediction Model}
\label{sec:pre_model}
As shown in the right box in Figure~\ref{fig:aggregation}, after embedding aggregation, we finally obtain one embedding vector for each \texttt{Smali} class. 
The last step is to apply the learned embedding to downstream tasks. 
Typically, this can be done by feeding the embeddings to another independent neural network that can be trained to a specific task.
In the original BERT, they only add one additional layer following the pre-trained model for each downstream task. 

In our approach, we design a simple model architecture with only three fully connected layers of neural network as the task-specific model and freeze the parameters of the pre-trained \toolname model when tuning for specific tasks.
The computational cost in training downstream tasks is significantly decreased by only tuning parameters of the task-specific model.
Specifically, the parameter number of the task-specific model is 10.4K, which is almost negligible compared with 459.35M of the pre-trained \toolname model.

The input of the task-specific model is the aggregated vector of class embedding, as shown in Figure~\ref{fig:aggregation}.
The task-specific model for malicious code localization predicts whether what a given class contains is malicious or not.
Similarly, for defect detection, the task-specific model predicts if a given class contains defective code.
{For component type classification, the task-specific model predicts to which component type the given class belongs.}
The details of each task-specific setup are presented in Section~\ref{sec:setup}.
\section{Study Design}
\label{sec:design}
In this section, we first overview the research questions to investigate in Section~\ref{sec:rqs}.
Then, details about dataset and empirical setup are presented in Section~\ref{sec:dataset} and Section~\ref{sec:setup}.
We provide evaluation results and answer the research questions in Section~\ref{sec:evalution}.

\subsection{Research Questions}
\label{sec:rqs}
In this paper, we consider the following six main research questions:
\newcommand{\rqone}{{Can \toolname accurately model Smali bytecode?}\xspace}
\newcommand{\rqtwo}{{How effective is the \toolname representation for the task of Malicious Code Localization?}\xspace}
\newcommand{\rqthree}{{How effective is the \toolname representation for the task of Defect Detection?}\xspace}
\newcommand{\rqfour}{{{How effective is the \toolname representation for the task of Component Type Classification?}}\xspace}
\newcommand{\rqfive}{{What are the impacts of different aggregation methods of instruction embeddings?}\xspace}
\newcommand{\rqsix}{{Can \toolname work with subsets of instructions?}\xspace}
\begin{itemize}[leftmargin=*]
    \item \textbf{RQ1:} \rqone
    \item \textbf{RQ2:} \rqtwo
    \item \textbf{RQ3:} \rqthree
    \item \textbf{RQ4:} \rqfour
    \item \textbf{RQ5:} \rqfive
    \item \textbf{RQ6:} \rqsix
\end{itemize}

\subsection{Dataset}
\label{sec:dataset}
In this work, we rely on four different datasets that we present in this section.

\subsubsection{Dataset for Pre-training}
\label{sec:pre_data}
With \toolname,  we target a general-purpose representation for various Android analysis tasks.
To obtain a representative sampling of the diverse landscape of android apps, we opted to leverage the dataset of a recent work in Android malware detection. 
{One thousand} apps---malware or benign---are randomly selected from the dataset used to evaluate DexRay~\cite{daoudi2021dexray}, a work that collected more than \num{158000} apps from the AndroZoo dataset~\cite{allix2016androzoo}. 

Class de-duplication was performed in order to include as much diversity as possible without letting the total number of instructions explode.
After removing duplicate classes, our selection of APKs results in over \textbf{35 million} \texttt{Smali} instructions, {from which we obtained \textbf{556 million tokens}. This is comparable to the scale of BooksCorpus~\cite{zhu2015aligning}, one of the pretraining datasets used in the original BERT, which consists of 800 million tokens.}

{Despite the pre-training dataset of DexBERT being smaller than the original BERT, its sufficiency is supported by two factors.
Firstly, \texttt{Smali}, being an assembly language, possesses a simpler structure and a significantly smaller set of tokens compared to natural languages and high-level programming languages, implying that a smaller dataset is enough to capture its essential features.
Secondly, the efficiency of the dataset is evaluated based on DexBERT's performance in downstream tasks which use APKs from distinct sources than the pre-training dataset. 
The superior performance of DexBERT over baseline models in these tasks confirms the adequacy of the pre-training dataset.}

Based on the \texttt{Smali} instructions in the dataset, we generated a WordPiece~\cite{wu2016google} vocabulary of \num{10000} tokens for \toolname, which is only one-third the size of the original BERT vocabulary. 
{The WordPiece model employs a subword tokenization method to manage extensive vocabularies and handle rare and unknown words. 
It breaks down words into smaller units, effectively addressing out-of-vocabulary words.}

\subsubsection{Dataset for Malicious Code Localization}
\label{sec:mcl_data}
\textbf{RQ2} deals with Malicious Code Localization, \textit{i.e.}, finding what part(s) of a given malware contains malicious code. 
At least two existing works have tackled this challenging problem for Android Malware and thus have acquired a suitable dataset with ground-truth labels. 
In Mystique~\cite{meng2016mystique}, Meng~\textit{et al.} constructed a dataset of \num{10000} auto-generated malware, with malicious/benign labels for each class.
However, almost all of the code in these generated APKs is either malicious or from commonly used libraries (such as \texttt{android.support}), and thus may not be representative of existing apps, nor of the diversity of Android apps.
More recently, in MKLDroid~\cite{narayanan2018multi}, Narayanan~\textit{et al.} randomly selected \num{3000} apps from the Mystique dataset and \emph{piggybacked} the malicious parts into existing, real-world benign apps from Google Play, resulting in a dataset they named MYST. 
Although still a little far from the real-world scenario, the repackaged malware in the MYST dataset contains both malicious and benign classes, which can support a class-level malicious code localization task. We thus decide to rely on the MYST dataset to conduct our experiments related to RQ2.
Note that, to the best of our knowledge, no fully labeled dataset of real-world malware exists for the task of malicious code localization for Android.
Note also that in our work, we choose MKLDroid as a baseline work, enabling us to directly compare \toolname against MKLDroid.

{Despite the challenges in acquiring labeled real-world malware, we remained determined to evaluate \toolname's performance in real-world scenarios. Ultimately, we succeeded in constructing a dataset that, albeit not extensive, contains labeled real-world malicious classes, thus broadening our evaluation scope. 
Specifically, we found 46 apps in the Difuzer~\cite{samhi2022difuzer} dataset, where the locations of a specific malicious behavior, namely the logic bomb, have been manually labeled.
This allows us to obtain labels of malicious classes and thereby assess DexBERT's ability to localize malicious code in real-world applications.
In addition, the authors of Difuzer provided more apps from their subsequent work, and we were able to successfully download and process 88 apks in total.}

{Given that each APK in the Difuzer dataset only has one class labeled as containing a logic bomb, and the malicious or benign nature of other classes is unknown, we are only able to utilize 88 malicious classes from these 88 real-world APKs for our extended evaluation. To facilitate a more comprehensive evaluation process, we constructed a dataset with additional APKs.
The training set comprises three sources: 50 Difuzer APKs with logic bombs, 50 benign APKs from the DexRay dataset~\cite{daoudi2021dexray}, and 100 APKs from MYST dataset to augment the dataset size. 
Please note that we selectively choose a portion of the benign classes at random to prevent a significant data imbalance, as the initial number of benign classes is much larger than that of malicious classes. As a result, we acquired 1929 benign classes and 425 malicious classes, including 50 from Difuzer APKs, for fine-tuning the classifier.
The evaluation set, aimed at testing DexBERT on real-world APKs, consists solely of the remaining 38 APKs with logic bombs from Difuzer and 50 benign APKs from DexRay. We ended up with 95 benign classes (almost two per benign apks) and 38 malicious classes containing logic bombs.
Please be aware that all the aforementioned designs aim to make the constructed data suitable for deep learning model training, while ensuring that there is no overlap between the training and evaluation sets.}

\subsubsection{Dataset for App Defect Detection}
\label{sec:defect_data}
As another important class-level Android analysis task, app defect detection is the subject of \textbf{RQ3}.
Dong~\textit{et al.} proposed \texttt{smali2vec} as a deep neural network based approach to detect application defects and released a dataset containing more than 92K Smali class files collected from ten Android app projects in over fifty versions.
For the convenience of labeling, they collected these APKs from GitHub based on three project selection criteria: 
1) the number of versions is greater than 20; 
2) the package size is greater than 500 KB; 
and 3) a large number of commits and of contributors.
The defective \texttt{Smali} files are located and labeled with  Checkmarx~\cite{checkmarx}, a widely used commercial static source code analysis tool.
Finally, each \texttt{Smali} class in the dataset has a label indicating whether it is defective or not.
We choose \texttt{smali2vec} as our dataset for app defect detection to enable comparison with their \texttt{smali2vec} approach built specifically for defect detection.

\subsubsection{Dataset for Component Type Classification}
\label{sec:component_data}
{To further evaluate the universality of \toolname, we introduced a third class-level task called Component Type Classification. This task, distinctly different from the previous two, is designed to provide a comprehensive assessment of \toolname's applicability across various scenarios.
In the Android framework, four primary components exist, namely Activities, Services, Broadcast Receivers, and Content Providers. 
These are fundamental building blocks of an Android application and are declared in an application’s manifest file (\texttt{AndroidManifest.xml}). 
Therefore, we can readily obtain labels for these four types of component classes and formulate a high-quality dataset for this task. 
Note that this task was designed solely to demonstrate the universality of DexBERT; It is selected due to the different nature of this task from other two downstream tasks and the ease and speed with which ground truth can be obtained. 
We randomly selected 1000 real-world APKs from the AndroZoo repository~\cite{allix2016androzoo}, from which we extracted 3406 component classes with accurate labels.
We used 75\% of this data for training and the remaining 25\% for testing. }

\subsection{Empirical Setup}
\label{sec:setup}

\subsubsection{Pre-training}
\label{sec:setup_pre}

Based on the typical BERT~\cite{bert} design, we simplify the model architecture of  \toolname to a certain extent to reduce the computational cost.
Indeed, while the dimension of intermediate layers in the position-wise feed-forward network was originally defined as $H \times 4$, where $H$ is set to 768 by default, as mentioned in Section~\ref{sec:ae}, we  reduce this dimension to $H \times 3$, \textit{i.e.}, from \num{3072} to \num{2304}.
The number of hidden layers and heads in the multi-head attention layers are set to 8 instead of 12. 
With these simplifications, the number of floating-point operations (FLOPs, indicating the computational complexity of the model) is reduced by 43.9\%, from 44.05G to 24.72G.
Meanwhile, the number of model parameters is only decreased by 7.7\%, from 497.45M to 459.35M.
Thus we keep as many as possible of the model parameters while reducing the computational cost, with the goal of preserving the learning ability of the model as much as possible.

The batch size is set to 72, and the learning rate is set to $1e^{-4}$.
Following the reference implementation\footnote{\url{https://github.com/dhlee347/pytorchic-bert}} we leveraged, we select the Adam optimizer~\cite{kingma2014adam}.
We adopt the Cross-Entropy loss function\footnote{\url{https://pytorch.org/docs/stable/generated/torch.nn.CrossEntropyLoss.html}} for both the \emph{masked words prediction} task and the \emph{next sentence prediction} task, and the Mean Squared Error (MSE) loss function\footnote{\url{https://pytorch.org/docs/stable/generated/torch.nn.MSELoss.html}} for the Auto-Encoder task.
Particularly, the MSE Loss is a criterion that measures the squared L2 norm between each element in the input x and target y.
Thus, the loss value of this task could also be regarded as an evaluation metric for the Auto-Encoder task.

{The pre-training of DexBERT on 556 million tokens for 40 epochs took about 10 days on 2 Tesla V100 GPUs (each with 32G memory). 
However, it is important to note that DexBERT users do not need to pre-train the model from scratch. 
They can directly use the pre-trained model we provide for their own Android analysis tasks.}

\subsubsection{Malicious Code Localization}
As discussed, malicious code localization is a difficult and under-explored problem.
Therefore, there are few widely recognized approaches and benchmark datasets available. 
To the best of our knowledge, Narayanan~\textit{et al.} provided the first well-labeled dataset that comes close to real-world practical needs, to evaluate their multi-view context-aware approach MKLDroid~\cite{narayanan2018multi} for malicious code localization.

In order to fairly compare with this baseline work, we follow the same validation strategy, \textit{i.e.}, \num{2000} APKs for fine-tuning and the remaining \num{1000} APKs for evaluation. 
We use 3-fold cross validation to ensure the reliability of the results.
The prediction model consists of 3 fully connected layers, of with 128, 64, and 32 neurons, respectively.
The output layer consists of two neurons that are used to predict the probabilities for a class being either malicious or benign.
We leverage the best aggregation method (\textit{i.e.}, element-wise addition) found in \textbf{RQ5}.
When fine-tuning for this task, we use the Cross-Entropy loss function and the Adam optimizer.
With a batch size of 256, we train the prediction model for \num{40} epochs.
For evaluation metrics, we adopt Precision, Recall, False Positive Rate (FPR), and False Negative Rate (FNR), the same as the baseline work MKLDroid for easy comparison. 
We further report F1-Scores as the overall metric.


\subsubsection{App Defect Detection}
Similar to malicious code localization, app defect detection is a relatively new challenge, only addressed by a small number of works in the literature.
Particularly, Dong~\textit{et al.} proposed a DNN-based approach \texttt{smali2vec}~\cite{dong2018defect} targeting Android applications and released a benchmark dataset containing more than 92K Smali class files.
{The classifier of \texttt{smali2vec} has 810,600 weight parameters across 10 layers, each of which has 300 neurons.
In contrast, our model only comprises a total of 10,304 weight parameters spanning three simple layers with sizes of 128, 64, and 32.}

In this work, we follow their \emph{Within-Project Defect Prediction} (WPDP) strategy using 5-fold cross-validation to compare their approach with ours. 
They provide an AUC score for each project, and report their mean value as the final evaluation metric. 
In addition, we also report the weighted average score to cater to the significant size variations among projects.
For this task, we adopt the same model architecture, aggregation method, loss function, and training strategy as for the malicious code localization task.

\subsubsection{Component Type Classification}
{The majority of the empirical settings for the Component Type Classification task are identical to those used in Malicious Code Localization, with the exception of the number of neurons in the output layer of the prediction model. 
In this task, the classifier contains four neurons, instead of two, to output the probabilities for the four different component types.}

\begin{figure*}[h]
\centering
\includegraphics[keepaspectratio,width=0.95\linewidth]{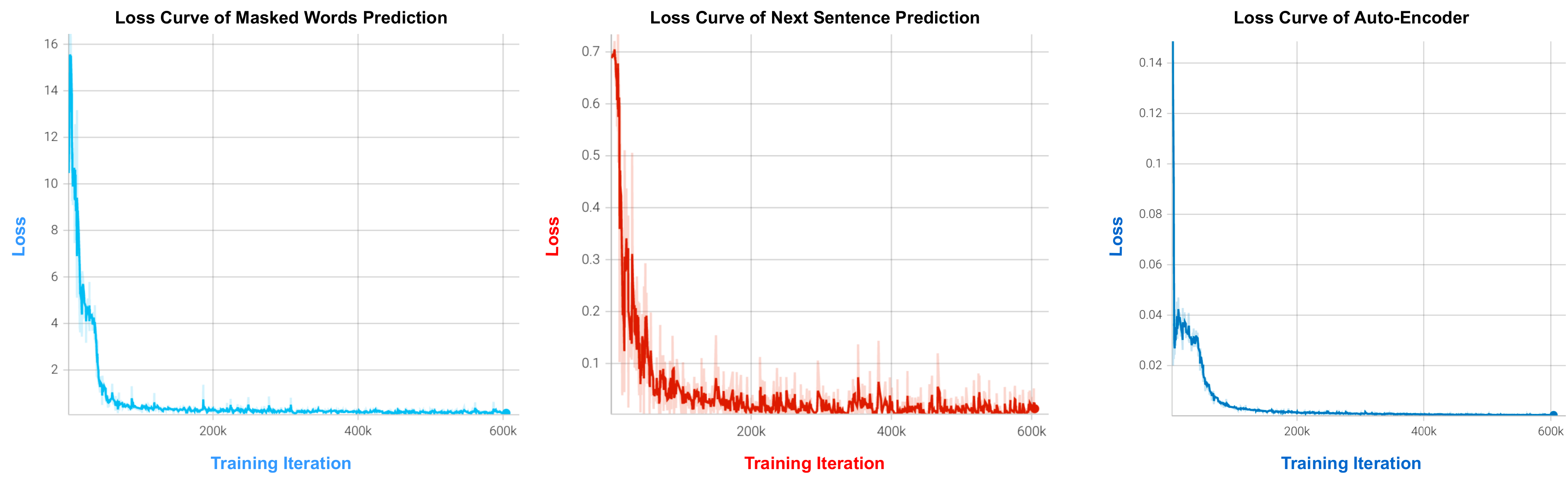}
\caption{Loss Curves of Three Pre-training Tasks. The X axis represents the training iteration index and Y axis represents the loss value.
}
\label{fig:pre_loss}
\end{figure*}

\section{Experimental Results}
\label{sec:evalution}

In this section, we present the evaluation results of \toolname, and we answer our five research questions.

\subsection{RQ1: \rqone}


With this first RQ, we assess whether or not \toolname is able to learn from \texttt{Smali} bytecode and build an accurate model of \texttt{Smali} bytecode as used in Android apps.
To that end, we report the loss curves of the three pre-training tasks, presented in Sections~\ref{sec:pretraining} and \ref{sec:ae}. 
These loss curves are shown in Figure~\ref{fig:pre_loss}, where the X axis represents the training iterations (i.e., batches). 

Two elements allow us to confidently conclude that \toolname indeed can learn an accurate model of \texttt{Smali} bytecode. 
First, the loss for all three pre-training tasks rapidly drops and is already very low after being fed just a small portion of our pre-training dataset, suggesting that our dataset is more than large enough for our purpose.
Second, continuing the pre-training process results in even lower loss and does not generate random fluctuations, suggesting that the model learned is not contradicted by new pieces of \texttt{Smali} bytecode and indeed converges. 

As mentioned in Section~\ref{sec:setup_pre}, the MSE loss could also be regarded as an evaluation metric, and its loss value in the third curve in Figure~\ref{fig:pre_loss} approaches zero late in the training process, indicating that the Auto-Encoder is able to reconstruct the given input vector of \toolname embedding with minimal error.
Therefore, the output vector of Encoder (\textit{i.e.}, Hidden Embedding in Figure~\ref{fig:detail_androbert}) preserves the key information of the original \toolname representation, which is required for the representation reconstruction by Decoder.

\begin{table}[h]
  \caption{Evaluation of Pre-training Tasks. The masked words prediction task is evaluated on \num{2037400} tokens and the next sentence prediction task is evaluated on \num{101870} instruction pairs.}
  \label{tab:pre_task}
  \begin{center}
  \begin{tabular}{|c|c|c|}
  \hline
  Task                     & Accuracy & \# of Samples            \\ \hline
  Masked Words Prediction  & 95.30\%  & \num{2037400}            \\ \hline
  Next Sentence Prediction & 99.35\%  & \num{101870} \\ \hline
  \end{tabular}
  \end{center}
\end{table}

In order to further evaluate the other two pre-training tasks, we created an \textbf{evaluation set} containing \num{2037400} masked tokens for the masked words prediction task and \num{101870} instruction pairs for next sentence prediction task.
We calculate accuracy on the evaluation set as a metric to further evaluate the performance of \toolname on these two tasks.

{Given the initial imbalanced distribution of each token and the randomness of our selection process, the distribution of masked tokens was imbalanced.
Common characters, strings, or variables such as slash "/" (15.47\%), comma ``,'' (6.32\%), ``v0'' (1.83\%), ``object'' (1.73\%), and ``lcom'' (1.67\%) were most frequently masked. Nevertheless, less common characters ($<$ 1.00\%) still accounted for 42.69\% of the masked tokens.
As shown in Table~\ref{tab:pre_task}, with an accuracy of 95.30\% on over two million predictions, we believe DexBERT's performance in the MLM task is robust.
Regarding the NSP task, the distribution of positive and negative samples was well balanced; positive samples constituted 49.81\% of the total samples, and negative samples made up the remaining 50.19\%. DexBERT achieved a near-perfect accuracy of 99.35\% on over 100K instruction pairs. 
This suggests that DexBERT was able to learn accurate features for the NSP task.}
The high accuracy of these two tasks demonstrates that the learned representations contain key features of the input instruction sequences leading to correct predictions.

\begin{tcolorbox}[enhanced,width=\linewidth,drop fuzzy shadow southwest,
    colback=gray!10,boxsep=0mm]
\textbf{RQ1 Answer:} \toolname can learn an accurate model of Smali bytecode.
\end{tcolorbox}

\subsection{RQ2: \rqtwo}
In this section, we investigate the performance of \toolname on the malicious code localization task and compare it with the MKLDroid baseline work~\cite{narayanan2018multi} on their evaluation dataset.
Following their experimental setup, we fine-tune \toolname to output, for each class of a given app, a \emph{maliciousness score}, or \emph{m-score} for short.
MKLDroid was evaluated with a beam search strategy, with a width of 10. 
In Table~\ref{tab:mcd_task}, we show MKLDroid performance metrics as reported in the MKLDroid paper~\cite{narayanan2018multi}, followed by the performance metrics of \toolname (Line \emph{\toolname-m}), also computed with a beam search of width 10. 


Additionally, we perform an experiment where we evaluate \toolname without beam search, where each class is predicted as malicious if the m-score is above a threshold of \num{0.5} or benign otherwise. 
The performance metrics for this experience are reported in the last line of Table~\ref{tab:mcd_task}. 
In effect, when evaluated in the same conditions as MKLDroid, \toolname significantly outperforms MKLDroid {with an F1 score 0.9981 on the MYST dataset}. 
Therefore, \linebreak \toolname does not need beam-search at all and achieves excellent performance when classifying each class independently. 
{Furthermore, we also include \texttt{smali2vec}~\cite{dong2018defect} as an additional baseline, which, although it achieves fairly good performance, fails to outperform our proposed \toolname.}


\begin{table}[htp]
\caption{Performance of Malicious Code Localization {on the MYST dataset.}}
\label{tab:mcd_task}
\begin{tabular}{|cccccc|}
\hline
\multicolumn{1}{|c|}{Approach} & \multicolumn{1}{c|}{F1-Score} & \multicolumn{1}{c|}{Precision} & \multicolumn{1}{c|}{Recall} & \multicolumn{1}{c|}{FNR} & {FPR} \\ \hline
\multicolumn{1}{|c|}{MKLDroid} & \multicolumn{1}{c|}{0.2488}   & \multicolumn{1}{c|}{0.1434}    & \multicolumn{1}{c|}{0.9400} & \multicolumn{1}{c|}{0.0500} & 0.1700 \\ \hline
\multicolumn{1}{|c|}{smali2vec} & \multicolumn{1}{c|}{0.9916}   & \multicolumn{1}{c|}{0.9880}    & \multicolumn{1}{c|}{0.9954} & \multicolumn{1}{c|}{0.0046} & 0.0046 \\ \hline
\multicolumn{1}{|c|}{\toolname-m} & \multicolumn{1}{c|}{0.5749}   & \multicolumn{1}{c|}{0.4034}    & \multicolumn{1}{c|}{\textbf{1.0000}} & \multicolumn{1}{c|}{\textbf{0.0000}} & 0.4847 \\ \hline
\multicolumn{1}{|c|}{\toolname} & \multicolumn{1}{c|}{\textbf{0.9981}} & \multicolumn{1}{c|}{\textbf{0.9983}} & \multicolumn{1}{c|}{0.9979} & \multicolumn{1}{c|}{0.0021} & \textbf{0.0006} \\ \hline
\end{tabular}
\end{table}

{As noted in Section~\ref{sec:mcl_data}, we expanded our evaluation of \toolname on real-world Android applications. Employing our dataset constructed from Difuzer apps{, i.e., Difuzer Extension dataset}, \toolname achieved a notable F1 score of 0.9048 in identifying malicious classes within real-world APKs. Further, it achieved a commendable F1 score of 0.9560 in predicting benign classes, thereby eliminating our concerns that data imbalance might negatively impact the evaluation.}


\begin{tcolorbox}[enhanced,width=\linewidth,drop fuzzy shadow southwest,
    colback=gray!10,boxsep=0mm]
\textbf{RQ2 Answer:} \toolname significantly outperforms MKLDroid on the task of malicious code localization when evaluated in the same conditions. 
In addition, \toolname can achieve vastly superior results when classifying each class independently. 
{Furthermore, \toolname also show its potential on localizing malicious classes within real-world Android apps.}
\end{tcolorbox}

\subsection{RQ3: \rqthree}

\begin{table*}[]
\caption{Performance of App Defect Detection}
\label{tab:defect_task}
\resizebox{\linewidth}{!}{
\begin{tabular}{|*{13}{c|}} 
\cline{1-13} 
\hline
\multicolumn{1}{|c|}{Project} & \tiny AnkiDroid & \tiny BankDroid & \tiny BoardGame & \tiny Chess & \tiny ConnectBot & \tiny Andlytics & \tiny FBreader &\tiny  K9Mail & \tiny Wikipedia & \tiny Yaaic & Average &  Weighted Average \\
\multicolumn{1}{|c|}{  \# of classes} &    \num{14767} &   \num{12372} &   \num{1634}&   \num{5005} &   \num{3865} &   \num{5305} &   \num{9883} &   \num{11857} &   \num{18883} &   \num{974} &   Score &   AUC Score  \\ \hline
\multicolumn{1}{|c|}{  \texttt{smali2vec}} &
  \multicolumn{1}{c|}{  0.7914} &
  \multicolumn{1}{c|}{  0.7967} &
  \multicolumn{1}{c|}{  \textbf{0.8887}} &
  \multicolumn{1}{c|}{  0.8481} &
  \multicolumn{1}{c|}{  \textbf{0.9516}} &
  \multicolumn{1}{c|}{  0.834} &
  \multicolumn{1}{c|}{  0.8932} &
  \multicolumn{1}{c|}{  0.7655} &
  \multicolumn{1}{c|}{  \textbf{0.8922}} &
  \multicolumn{1}{c|}{  \textbf{0.9371}} &
  \multicolumn{1}{c|}{  0.8598} &
    0.8399 \\ \hline
\multicolumn{1}{|c|}{  \toolname} &
  \multicolumn{1}{c|}{\textbf{  0.9572}} &
  \multicolumn{1}{c|}{  \textbf{0.9363}} &
  \multicolumn{1}{c|}{  0.7691} &
  \multicolumn{1}{c|}{  \textbf{0.9125}} &
  \multicolumn{1}{c|}{  0.8517} &
  \multicolumn{1}{c|}{  \textbf{0.9248}} &
  \multicolumn{1}{c|}{  \textbf{0.9378}} &
  \multicolumn{1}{c|}{  \textbf{0.8674}} &
  \multicolumn{1}{c|}{  0.8587} &
  \multicolumn{1}{c|}{  0.8764} &
  \multicolumn{1}{c|}{\textbf{  0.8892}} &
   \textbf{0.9032} \\ \hline
\end{tabular}
}
\vspace{-3mm}
\end{table*}

In this section, we investigate the performance of \toolname on the task of app defect detection, and we compare it against the baseline work \texttt{smali2vec}~\cite{dong2018defect}.
The performance of \texttt{smali2vec}\footnote{as reported in the \texttt{smali2vec} paper~\cite{dong2018defect}} on 10 Android projects is shown in Table~\ref{tab:defect_task}, where the \texttt{\# of classes} represents the number of \texttt{Smali} classes in each project.

Using our \toolname representation, we fine-tune a model to predict the likelihood that a given class is defective.

As shown in Table~\ref{tab:defect_task}, \toolname outperforms \texttt{smali2vec} on 6 out of 10 projects and achieves a weighted AUC score of 90.32\%, which is a 6.33 percentage points improvement over \texttt{smali2vec}.

\begin{tcolorbox}[enhanced,width=\linewidth,drop fuzzy shadow southwest,
    colback=gray!10,boxsep=0mm]
\textbf{RQ3 Answer:} \toolname slightly outperforms \texttt{smali2vec} for the task of app defect detection.
\end{tcolorbox}

\subsection{RQ 4: \rqfour}
\label{sec:rq_4}
{In this section, we explore \toolname's performance on the Component Type Classification task, aiming to further examine its universal applicability across diverse application scenarios. We contrast \toolname's performance against other BERT-like models, specifically BERT~\cite{bert}, CodeBERT~\cite{CodeBERT}, and DexBERT without pre-training. Similar to the settings for the Malicious Code Localization task, we fine-tune a classifier to predict a given class's component type. Given that other BERT-like baselines lack an AutoEncoder module for further reduction of embedding dimensionality, we use the first state vector (size 768) of the embedding for all comparative experiments.}

{As Table~\ref{tab:component} illustrates, DexBERT excels in predicting all four types of component classes. On average, DexBERT's performance surpasses all baselines by a significant margin, exhibiting a roughly 20 percentage point increase in terms of F1 Score. This reiterates DexBERT's effectiveness in representing Smali instructions in Android and, crucially, validates the universality of DexBERT.}

\begin{tcolorbox}[enhanced,width=\linewidth,drop fuzzy shadow southwest,
    colback=gray!10,boxsep=0mm]
\textbf{RQ4 Answer:} {\toolname significantly outperforms baselines for the task of component type classification which differs from first two tasks that focused on security-related properties, demonstrating its versatility across various application scenarios.}
\end{tcolorbox}

\begin{table}[]
\caption{Comparison of F1 Score Among Various BERT-like Baselines for Four Component Classes.}
\label{tab:component}
  \begin{adjustbox}{width=\columnwidth,center}
\begin{tabular}{|c|c|c|c|c|c|}
\hline
\multicolumn{1}{|c|}{Method} & \multicolumn{1}{c|}{Activity} & \multicolumn{1}{c|}{Service} & \multicolumn{1}{c|}{BroadcastReceiver} & \multicolumn{1}{c|}{ContentProvider} & \multicolumn{1}{c|}{Average} \\ \hline
\multicolumn{1}{|c|}{BERT} & \multicolumn{1}{c|}{0.8272} & \multicolumn{1}{c|}{0.7642} & \multicolumn{1}{c|}{0.5673} & \multicolumn{1}{c|}{0.9091} & \multicolumn{1}{c|}{0.7669} \\ \hline
\multicolumn{1}{|c|}{CodeBERT} & \multicolumn{1}{c|}{0.917} & \multicolumn{1}{c|}{0.5381} & \multicolumn{1}{c|}{0.8756} & \multicolumn{1}{c|}{0.8468} & \multicolumn{1}{c|}{0.7943} \\ \hline
\multicolumn{1}{|c|}{DexBERT(woPT)} & \multicolumn{1}{c|}{0.7402} & \multicolumn{1}{c|}{0.5850} & \multicolumn{1}{c|}{0.7660} & \multicolumn{1}{c|}{0.8947} & \multicolumn{1}{c|}{0.7465} \\ \hline
\multicolumn{1}{|c|}{DexBERT} & \multicolumn{1}{c|}{\textbf{0.9780}} & \multicolumn{1}{c|}{\textbf{0.9117}} & \multicolumn{1}{c|}{\textbf{0.9600}} & \multicolumn{1}{c|}{\textbf{0.9756}} & \multicolumn{1}{c|}{\textbf{0.9563}} \\ \hline
\end{tabular}%
\end{adjustbox}%
\end{table}%

\subsection{RQ 5: \rqfive}
\label{sec:rq_aggregate}

\begin{table}[]
\caption{Comparison of Different Aggregation Methods on Three Downstream Tasks: Malicious Code Localization (MCL), Defect Detection (DD), and Component Type Classification (CTC)}
  \begin{adjustbox}{width=\columnwidth,center}
\label{tab:aggregation}
\begin{tabular}{|c|c|c|c|}
\hline
\multicolumn{1}{|c|}{Method} & \multicolumn{1}{c|}{MCL@F1 Score} & \multicolumn{1}{c|}{DD@AUC Score} & \multicolumn{1}{c|}{CTC@F1 Score} \\ \hline
\multicolumn{1}{|c|}{Addition} & \multicolumn{1}{c|}{\textbf{0.9989}} & \multicolumn{1}{c|}{\textbf{0.9064}} & \multicolumn{1}{c|}{\textbf{0.9563}} \\ \hline
\multicolumn{1}{|c|}{Random} & \multicolumn{1}{c|}{0.9982} & \multicolumn{1}{c|}{0.8553} & \multicolumn{1}{c|}{0.8898} \\ \hline
\multicolumn{1}{|c|}{Average} & \multicolumn{1}{c|}{0.9916} & \multicolumn{1}{c|}{0.8712} & \multicolumn{1}{c|}{0.9442} \\ \hline
\multicolumn{1}{|c|}{Concat\&Resize} & \multicolumn{1}{c|}{0.9979} & \multicolumn{1}{c|}{0.8508} & \multicolumn{1}{c|}{0.7491} \\ \hline
\end{tabular}%
\end{adjustbox}
\end{table}%

In this section, we investigate the impact of the different embedding aggregation methods (\textit{i.e.}, element-wise addition, {random selection}, averaging and concatenation \& bilinear resizing, cf. Section~\ref{sec:aggregation}).
{These techniques were initially leveraged to aggregate token embeddings in BERT~\cite{bert} and other deep learning approaches~\cite{daoudi2021dexray}. From our point of view,  the output features from BERT and the AutoEncoder of \toolname are essentially similar in nature. Each state vector of BERT embedding is a high-level abstract feature of the corresponding token. Similarly, each output vector of Auto-Encoder is a high-level abstract feature vector of the corresponding token sequence. Therefore, if it's plausible to aggregate token embeddings by addition or the other three techniques, it should also be plausible to aggregate sequence embeddings in a similar manner.}

We conduct comparative experiments based on three downstream tasks to evaluate to what extent \toolname is sensitive to the aggregation method.
As shown in Table~\ref{tab:aggregation}, on the malicious code localization task, performance metrics for all four methods are very close, with no significant differences between methods.
For the other two downstream tasks, we note that the differences are more significant than for the task of malicious code localization. 
Despite all four aggregation methods yielding an acceptable performance, \emph{addition} is the best performer, achieving the highest metric scores on both tasks.


\begin{tcolorbox}[enhanced,width=\linewidth,drop fuzzy shadow southwest,
    colback=gray!10,boxsep=0mm]
\textbf{RQ5 Answer:} All four proposed aggregation methods are effective on all  three downstream tasks. 
Element-wise addition achieves the best performance on both tasks.
\end{tcolorbox}

\subsection{RQ6: \rqsix}


\begin{table}[]
 \caption{Comparative Analysis of Full Instructions vs API Calls for Malicious Code Localization (MCL) and Component Type Classification (CTC). "Avg Time" means the average inference time per class.}
 \label{tab:comp_api}
  \begin{adjustbox}{width=\columnwidth,center}
\begin{tabular}{|ccccc|}
\hline

\multicolumn{1}{|c|}{\small Method} & \multicolumn{1}{c|}{\small MCL@F1 Score} & \multicolumn{1}{c|}{\small CTC@F1 Score} & \multicolumn{1}{c|}{\small Avg Time} \\ \hline
\multicolumn{1}{|c|}{\small Full Instructions} & \multicolumn{1}{c|}{\textbf{0.9981}} & \multicolumn{1}{c|}{\textbf{0.9563}} & \multicolumn{1}{c|}{0.00768s} \\ \hline
\multicolumn{1}{|c|}{\small API Call} & \multicolumn{1}{c|}{0.9932}   & \multicolumn{1}{c|}{0.8779} & \multicolumn{1}{c|}{\textbf{0.00073s}}    \\ \hline
\end{tabular}%
\end{adjustbox}%
\end{table}%

In the previous RQs, we demonstrated the effectiveness of \toolname when using the entire \texttt{Smali} bytecode. 
Representing \texttt{Smali} bytecode with \toolname can be computationally expensive, given the very large number of instructions an app (or even a class) can contain.
With this RQ, we investigate the ability of \toolname to work with subsets of instructions, hence reducing the number of pieces of code to represent and reducing the need for aggregation.

We postulate that API invocations are the instructions that carry the most semantics information, and thus conduct an experiment where we pre-filter the flow of \texttt{Smali} bytecode to keep only API calls. 
{Based on the statistics on our pre-training dataset, API instructions constitute approximately 17.27\% of the total instructions.}
We re-use the pre-trained model built with the complete flow of instructions (Cf. RQ~1), but we fine-tune a dedicated model with filtered instructions only. 
Since many classes in some projects in the dataset for defect detection do not have API calls {(some concrete examples are included in appendix)}, which would result in empty representations, we only consider the tasks of malicious code localization {and Component Type Classification} here.

As shown in Table~\ref{tab:comp_api}, while the performance of \toolname is slightly higher with all instructions, \toolname still performs very well with API calls only. 
Computationally, however, working with only API calls is one order of magnitude faster.
{As for the total execution time, we take the evaluation of malicious code localization task as an example here.
With full instructions, we require approximately 1.9 hours to generate the DexBERT features for all 911,724 classes. 
Conversely, with only API calls, we need only about 11 minutes to generate all feature vectors.}

\begin{tcolorbox}[enhanced,width=\linewidth,drop fuzzy shadow southwest,
    colback=gray!10,boxsep=0mm]
\textbf{RQ6 Answer:} When compatible with the downstream task, \toolname is also very effective and fast when considering API calls only. 
\end{tcolorbox}
\begin{table}[]
\centering
 \caption{Ablation Study on the Impact of DexBERT Embedding Size}
 \label{tab:AE_Ablation}
  \begin{adjustbox}{width=\columnwidth,center}
\begin{tabular}{|c|c|c|c|c|}
\hline
\multicolumn{1}{|c|}{\small Size} & \multicolumn{1}{c|}{\small MCL@F1 Score} & \multicolumn{1}{c|}{\small DD@AUC Score} & \multicolumn{1}{c|}{\small {DD@F1 Score}} & \multicolumn{1}{c|}{\small CTC@F1 Score} \\ \hline
\multicolumn{1}{|c|}{\small 768} & \multicolumn{1}{c|}{0.9999} & \multicolumn{1}{c|}{0.9699} & \multicolumn{1}{c|}{{0.8887}} & \multicolumn{1}{c|}{0.9563} \\ \hline
\multicolumn{1}{|c|}{\small 256} & \multicolumn{1}{c|}{0.9995}   & \multicolumn{1}{c|}{0.9336}  & \multicolumn{1}{c|}{{0.8029}}  & \multicolumn{1}{c|}{0.9246} \\ \hline
\multicolumn{1}{|c|}{\small 128} & \multicolumn{1}{c|}{0.9981}   & \multicolumn{1}{c|}{0.9032}  & \multicolumn{1}{c|}{{0.7542}}  & \multicolumn{1}{c|}{0.9202} \\ \hline
\multicolumn{1}{|c|}{\small 64} & \multicolumn{1}{c|}{0.9813}   & \multicolumn{1}{c|}{0.8472}  & \multicolumn{1}{c|}{{0.6693}}  & \multicolumn{1}{c|}{0.9007} \\ \hline
\end{tabular}%
\end{adjustbox}%
\end{table}%

\begin{table}[]
\centering
 \caption{Comparison of F1 Scores among Various BERT-Like Baselines for Three Tasks: Malicious Code Localization (MCL), Defect Detection (DD), and Component Type Classification (CTC). DexBERT(woPT) indicates DexBERT without Pre-training.}
 \label{tab:bert_baseline}
\begin{tabular}{|c|c|c|c|}
\hline
\multicolumn{1}{|c|}{\small Models} & \multicolumn{1}{c|}{\small MCL} & \multicolumn{1}{c|}{\small DD} & \multicolumn{1}{c|}{\small CC} \\ \hline
\multicolumn{1}{|c|}{\small BERT} & \multicolumn{1}{c|}{0.9182} & \multicolumn{1}{c|}{0.66851} & \multicolumn{1}{c|}{0.7669} \\ \hline
\multicolumn{1}{|c|}{\small CodeBERT} & \multicolumn{1}{c|}{0.9985}   & \multicolumn{1}{c|}{0.64775}    & \multicolumn{1}{c|}{0.7943} \\ \hline
\multicolumn{1}{|c|}{\small DexBERT(wo-PreT)} & \multicolumn{1}{c|}{0.9961}   & \multicolumn{1}{c|}{0.74381}    & \multicolumn{1}{c|}{0.3028} \\ \hline
\multicolumn{1}{|c|}{\small DexBERT} & \multicolumn{1}{c|}{\textbf{0.9999}}   & \multicolumn{1}{c|}{\textbf{0.8725}}    & \multicolumn{1}{c|}{\textbf{0.9563}} \\ \hline
\end{tabular}%
\end{table}%

\section{Discussion}
\label{sec:discussion}
In this section, {we begin with an ablation study examining the impact of \toolname's embedding size on downstream tasks, and an ablation study assessing the effectiveness of the two pre-training tasks. Following that, we compare the performance of various BERT-like baselines across three different downstream tasks.}
Then, we share some insights about the proposed \toolname for Android representation.
Next, we discuss some potential threats to validity of the proposed approach.
Finally, we discuss some future works which are worth studying next.

\subsection{Ablation Study on \toolname Embedding Size}
\label{sec:AE_Ablation}
{As detailed in Section~\ref{sec:ae}, Android application scenarios require a smaller embedding due to the considerably larger token quantities compared to typical textual documents and code files.
To find a reasonable trade-off between model computation cost and performance, we conducted an ablation study exploring the impact of \toolname embedding size on the three downstream tasks.
The experiments contain three different sizes for the hidden embedding of the AutoEncoder (AE), specifically 256, 128, and 64. Additionally, we evaluated the performance by directly utilizing the first state vector of the raw DexBERT embedding, which has a size of 768, without applying any dimension reduction from the AutoEncoder.} 

{Table~\ref{tab:AE_Ablation} reveals that in the task of Malicious Code Localization, a decrease in vector size does not lead to a significant loss in the performance, until the size is reduced to 128. Hence, we concluded that 128 is the optimal size for this task.}

{As for the tasks of Defect Detection and Component Type Classification, the experimental results demonstrate that a larger embedding size resulted in a considerable improvement in performance. However, a size of 128 also offered a solid trade-off for these two tasks, supporting satisfactory performance with a metric score exceeding 0.9.}
{Please be aware that the choice to use the AUC score for defect detection was made in order to maintain consistency with the metric employed by the primary baseline for this task, namely, \texttt{smali2vec}~\cite{dong2018defect}.
However, to be consistent with the other two tasks, we have also included the F1 score for this task in Table~\ref{tab:AE_Ablation}.}

\subsection{Ablation Study on Pre-training Tasks}
{To better understand the pre-training process, we conducted an ablation to confirm the necessity and effectiveness of the two pre-training designs, i.e., MLM and NSP, on the final improvement of the model.}

{In models like BERT and DexBERT, multi-task learning with MLM and NSP is designed to generate universal features for a variety of tasks.
Removing either task diminishes the model's representational power. 
As demonstrated in Table~\ref{tab:pretrain_ablation}, while MLM alone can achieve relatively good performance, the combination of both pre-training tasks significantly improves the model's performance, reinforcing their mutual importance for capturing the smali bytecode structure and semantics effectively.
The results on all three downstream tasks, especially on Defect Detection (DD) and Component Type Classification (CTC) demonstrate the importance of both MLM and NSP pre-training tasks.}

\begin{table}[htbp]
  \centering
  \caption{Comparison of F1 scores on Three Downstream Tasks based on Different Pre-training Task Designs for : Malicious Code Localization (MCL), Defect Detection (DD), and Component Type Classification (CTC).}
    \begin{tabular}{|c|c|c|c|}
    \hline
    \multicolumn{1}{|c|}{Pre-training Designs} & \multicolumn{1}{c|}{MCL} & \multicolumn{1}{c|}{DD} & \multicolumn{1}{c|}{CTC} \\
    \hline
    \multicolumn{1}{|c|}{Only MLM} & 0.9987 & 0.8055 & 0.8827 \\
    \hline
    \multicolumn{1}{|c|}{Only NSP} & 0.6547 & 0.5331 & 0.5491 \\
    \hline
    \multicolumn{1}{|c|}{MLM\&NSP} & \textbf{0.9999} & \textbf{0.8725} & \textbf{0.9563} \\
    \hline
    \end{tabular}%
  \label{tab:pretrain_ablation}%
\end{table}%

\subsection{Comparative Study with other BERT-like Baselines}
\label{sec:bert-like_comparison}
{To better understand the necessity and effectiveness of the pre-training process on \texttt{Smali} code, in this section, we conduct a comparative study to assess the performance of existing BERT-like models that can be directly applied to all three Android downstream tasks without any technical barriers.
Specifically, the baselines include BERT~\cite{bert}, CodeBERT~\cite{CodeBERT}, and DexBERT without pre-training.
With the same reason in Section~\ref{sec:rq_4}, we use the first state vector (size 768) of the embedding for all comparative experiments.} 

{The outcomes are shown in Table~\ref{tab:bert_baseline}.     Interestingly, each baseline model performed remarkably well for Malicious Code Localization. This can be attributed to the fact that the dataset is artificially generated {by inserting malicious code into real-world apps}, resulting in a clear separation between positive and negative samples, making them easy to learn from. However, the pre-trained DexBERT model outperformed the baselines with an impressive F1 score of 0.9999, approaching perfection.}
    
{For the other two tasks, Defect Detection and Component Type Classification, where datasets were collected from real-world APKs, the pre-trained DexBERT clearly surpassed BERT and CodeBERT by approximately 20 percentage points on both tasks.
Furthermore, the performance of DexBERT without pre-training exhibited low performance and instability across the three tasks, which was somewhat expected as it lacked prior knowledge before being fine-tuned on downstream tasks. 
Overall, the comparative results shown in Table~\ref{tab:bert_baseline} strongly demonstrate the necessity and effectiveness of the DexBERT pre-training process on \texttt{Smali} code.}

\subsection{Insights}
\label{sec:insights}
In this work, we find that the popular NLP representation learning model BERT can be used for Android bytecode without much modification by regarding disassembled \texttt{Smali} instructions as natural language sentences. 
While it had already been shown that BERT-like models could be used for Source Code, our work shows it can also work directly with raw apps in the absence of original source code.

Still, there are some gaps between natural language and \texttt{Smali} code to mitigate.
In particular, NLP problems and Android analysis problems have significantly different application scenarios.
NLP problems are usually at the text snippet level, where the base unit is a (short) paragraph  (\textit{e.g.}, sentence translation or text classification). However, in Software Engineering and Security for Android applications, there are problems both at the class level and the whole app level, \textit{i.e.}, ranging from a few instructions to millions of instructions.
Therefore, embedding aggregation is required when applying instruction embeddings to Android analysis problems.

Besides, while this work is only evaluated on class-level tasks, it is expected to work at other levels (lower or higher).
There are no significant technical barriers for lower-level tasks (\textit{i.e.}, method level or statement level) except for the absence of well-labeled datasets. 
For higher-level tasks (\textit{i.e.}, APK level), further embedding aggregation would be required to support whole-application tasks.

\subsection{Threats to Validity}
\label{sec:threats}
Our experiments and conclusions may face threats to validity.
First, the MYST dataset used for malicious code localization is artificially created and, therefore may not be representative of the real-world landscape of apps.
{To mitigate this concern, we expanded our evaluation to include real-world apps from Difuzer~\cite{samhi2022difuzer}.}
{Moreover, the utility of our malicious code localization model in real-world scenarios could be further assessed by extending its application to a wider variety of malicious behaviours beyond logic bombs. This extension could be an avenue for future work, as it would necessitate significant manual analysis efforts.}

Second, the dataset for Android app defect detection was created several years ago. 
Android applications are constantly evolving over time.
How well \toolname performs on today's apps requires further validation.
Therefore, it may be necessary to create new datasets and conduct more comprehensive evaluations on Android app defect detection.

\subsection{Future Work}
\label{sec:future}
The proposed \toolname is validated on three class-level Android analysis tasks.
An important aspect would be to extend the range of tasks \toolname is evaluated on. 
We identified several tasks (such as malware detection, app clone detection, repackaging identification, \textit{etc.}) that are of interest to the Research community and that could benefit from our approach.

Besides, we showed \toolname representation may not always need the complete flow of \texttt{Smali} instructions.
However, we investigated only one filtering criterion. Other filtering approaches could be investigated to refine potential trade-offs between computational cost and effectiveness.

{Finally, the evaluation datasets for tasks such as malicious code localization and defect detection can be further enhanced by including more recent applications. Given the substantial efforts required to process and construct the new datasets, we plan to undertake this enhancement in a separate study in the future.}
\section{Related Work}
\label{sec:relatedwork}

This study lies at the intersection of the fields of Representation Learning and of Android app analysis.

\subsection{Representation Learning}
Recent successes in deep learning have attracted increased interest in applying deep learning techniques to learn representations of programming artifacts for a variety of software engineering tasks~\cite{allamanis2018survey, ernst2017natural, code2vec, CodeBERT}.

\subsubsection{Code Representation}
Code representation approaches aim to represent source code as feature vectors that contain the semantics and syntactic information of the source code. 
In general, code representations can be mainly categorized into sequence-based, tree-based, and graph-based representations. 
The tasks that rely on sequence-based representations consider source code as plain text and use traditional token-based methods to capture lexical information, such as clone detection~\cite{hua2020fcca}, vulnerability detection~\cite{xu2018vulnerability}, and code review~\cite{siow2020core}.
Tree-based representations capture features of source code by traversing the AST of the source code.
Code2Vec~\cite{code2vec} proposes a path-attention model to aggregate the set of AST paths into a vector.
TBCNN~\cite{tbcnn} learns code representations that capture structural information in the AST.
Tree-LSTM~\cite{TreeLSTM} employs LSTM in learning the network topology of the input tree structure of AST.
Graph-based representation approaches~\cite{wan2019multi,wang2020detecting} represent code as graphs that are associated with programs, such as control flow graph (CFG), control dependency graph (CDG), and data dependency graph (DDG). 

Inspired by the recent success of transformer-based language models like BERT~\cite{bert} and RoBERTa~\cite{roberta} in Natural Language Processing, Feng et al.~\cite{CodeBERT} proposed CodeBERT, which is pre-trained both on programming languages and natural languages.
Guo et al.~\cite{GraphCodeBERT} proposed GraphCodeBERT to advance CodeBERT by additionally considering data flow information in pre-training.




\subsubsection{Android app Representation}
Android app representations aim to represent an Android app into feature vectors for various tasks such as malware detection~\cite{arp2014drebin} and clone detection~\cite{chen2014achieving}.
Many works~\cite{xu2018cdgdroid,karbab2018maldozer,xiao2016identifying, canfora2015detecting, xiao2017back, singh2017dynamic, bhatia2017malware} relied on reverse engineering to extract information (features) from APKs and feed the extracted features into traditional ML-based and DL-based approaches to obtain Android representations~\cite{qiu2020survey}.
Static features such as permissions, API calls, and control flow graphs are widely used in prior works~\cite{li2018fine,booz2018tuning,grosse2017adversarial, pendlebury2019tesseract, xu2018cdgdroid,karbab2018maldozer,hou2016droiddelver} to generate Android representations.
There are several approaches where representation is based on dynamic features. 
For instance, several Android malware detectors~\cite{xiao2016identifying, canfora2015detecting, xiao2017back, singh2017dynamic, bhatia2017malware} leveraged system calls traces. 

The aforementioned features can be represented in different forms: the vectorized representation and the graph-based representation. 
Features such as permissions or API calls~\cite{li2018fine,booz2018tuning, grosse2017adversarial,pendlebury2019tesseract,naway2018using,naway2019android,lee2019seqdroid,he2018android}, raw~\cite{mclaughlin2017deep,Jerome2014} or processed~\cite{Allix2016:emse} opcode sequences, and dynamic behaviors~\cite{vinayakumar2018detecting,yuan2014droid} are mainly represented as vectors. 
Other graph-based features such as control flow graphs~\cite{Yang2014DroidMinerAM,Atici2016AndroidMA} and data flow graphs~\cite{Wei2014AmandroidAP} can be directly fed to DL models (\textit{e.g.}, Graph Convolutional Network~\cite{Bruna2014SpectralNA,gcn}) or embedded into vectors by graph embedding techniques (\textit{e.g.}, Graph2vec~\cite{graph2vec}).


\subsection{Android Analysis Tasks}
Android app analysis tasks can be conducted at different levels, such as APK-level (\textit{e.g.}, Android Malware Detection~\cite{arp2014drebin,hsien2018r2,daoudi2021dexray,sun2021android}, Android Repackaging Identification~\cite{shao2014towards,li2017simidroid,singh2021multi}), class-level or method-level (\textit{e.g.}, Malicious Code Localization, Android App Defect Detection).

\subsubsection{Android Malicious Code Localization}
\label{sec:related_mcl}

HsoMiner~\cite{pan2017dark} is an approach to discover HSO (Hidden Sensitive Operation) activities (\textit{e.g.}, stealing user's privacy). 
Li~\textit{et al.} proposed a tool called HookRanker~\cite{li2017locating} to automatically localize the malicious packages from piggybacked Android apps based on how malware behavior code is triggered. 
MKLDroid~\cite{narayanan2018multi}, proposed by Narayanan~\textit{et al.}, could be regarded as the first real malicious code localization approach. 
They consider multiple views of Android apps in a unified framework to detect malware.
MKLDroid assigns m-scores to every class, and the classes with the highest m-scores are considered malicious.  
Ma~\textit{et al.} proposed a deep learning based method called Droidetec~\cite{ma2020droidetec} for malware detection and malicious code localization by modeling an Android app as a natural language sequence.
MKLDroid and Doidetec could achieve a reasonable recall of malicious segments (classes or methods) by sacrificing precision. 
Recently, Wu~\textit{et al.} proposed a Graph Neural Network based approach~\cite{wu2021android} for Android malware detection and malicious code localization. 
Despite the help of manual checks, the obtained accuracy is still far from perfect. 
Among these related works above, only MKLDroid~\cite{narayanan2018multi} provides its replication package.


\subsubsection{Android App Defect Detection}
\label{sec:related_defect}
A software defect is an error or a bug caused by a programmer during the software design and development process. Early approaches for software defect detection~\cite{bowes2014dconfusion, perl2015vccfinder, scandariato2014predicting, wang2016automatically} were not easily adaptable for Android applications.

{Initial attempts at Android defect prediction~\cite{kaur2015investigation,kaur2016application} focused on extracting code and process metrics from mobile applications. Similarly, object-oriented metrics~\cite{ricky2016mobile,malhotra2016empirical} were employed to build defect prediction models. However, the feature engineering efforts in these approaches limited the verification of their methods' effectiveness to a small number of applications. Additionally, they relied on specific sets of code and process metrics, which might not be universally applicable to other APKs.} To address this limitation, Dong~\textit{et al.} proposed \texttt{smali2vec}~\cite{dong2018defect}, which automatically extracts features of \texttt{Smali} instructions and inputs them into a deep learning model to identify defective classes. They also provided a benchmark dataset for the community to advance the field of Android defect detection.

{Meanwhile, Just-in-Time (JIT) defect prediction~\cite{yan2020just,yan2020effort,zhao2021just} at the commit level has been developed, offering timely feedback for developers to detect defects early. However, class-level prediction methods remain necessary, as they help developers and testers prioritize their efforts by identifying the most defect-prone classes. Additionally, class-level prediction can be useful when a project has a low frequency of commits or uses a different version control system that makes commit-level prediction difficult. In this work, we focus on class-level Android defect detection.}



\section{Conclusion}
\label{sec:conclusion}
We propose a pre-trained representation learning model named \toolname aiming at solving various fine-grained Android analysis problems.
Based on Auto-Encoder, we design an aggregation method to overcome the input length limitation problem existing in the original BERT applications.
Freezing parameters of the pre-trained DexBERT model, the learned representation is able to be used directly on various class-level downstream tasks.
Comprehensive experimental results demonstrate its effectiveness on malicious code localization, Android application defect detection and component type classification, compared to baseline methods.

\section{Data Availability}
All artifacts of this study are available at:
    \url{https://github.com/Trustworthy-Software/DexBERT}

\section*{Acknowledgments}
This work was partially supported by the Fonds National de la Recherche (FNR), Luxembourg, under project REPROCESS C21/IS/16344458.

\bibliographystyle{IEEEtran}
\bibliography{references}



\vfill

\end{document}